\documentclass[aps,prl,twocolumn,unsortedaddress,superscriptaddress]{revtex4-1}

\usepackage{amsmath}
\usepackage{hyperref}
\usepackage{graphicx}
\usepackage{physics}
\usepackage{xfrac}
\usepackage{isotope}
\usepackage[all]{hypcap}

\hypersetup{colorlinks=true,citecolor=[rgb]{0,0,0.8},
linkcolor=[rgb]{0,0,0.8},urlcolor=[rgb]{0,0,0.8}}

\newcommand{\eu}[1]{\mathrm{e}^{#1}}

\usepackage[usenames]{color}
\usepackage[normalem]{ulem}

\begin{document}
\title{Is a Trineutron Resonance Lower in Energy than a Tetraneutron
Resonance?}

\author{S.~Gandolfi}
\email{stefano@lanl.gov}
\affiliation{Theoretical Division, Los Alamos National Laboratory,
Los Alamos, New Mexico 87545, USA}
\author{H.-W.~Hammer}
\email{hans-werner.hammer@physik.tu-darmstadt.de}
\affiliation{Institut f\"ur Kernphysik, 
Technische Universit\"at Darmstadt, 64289 Darmstadt, Germany}
\affiliation{ExtreMe Matter Institute EMMI, 
GSI Helmholtzzentrum f\"ur Schwerionenforschung GmbH, 64291 Darmstadt,
Germany}
\author{P.~Klos}
\email{pklos@theorie.ikp.physik.tu-darmstadt.de}
\affiliation{Institut f\"ur Kernphysik, 
Technische Universit\"at Darmstadt, 64289 Darmstadt, Germany}
\affiliation{ExtreMe Matter Institute EMMI, 
GSI Helmholtzzentrum f\"ur Schwerionenforschung GmbH, 64291 Darmstadt,
Germany}
\author{J.~E.~Lynn}
\email{joel.lynn@physik.tu-darmstadt.de}
\affiliation{Institut f\"ur Kernphysik, 
Technische Universit\"at Darmstadt, 64289 Darmstadt, Germany}
\affiliation{ExtreMe Matter Institute EMMI, 
GSI Helmholtzzentrum f\"ur Schwerionenforschung GmbH, 64291 Darmstadt,
Germany}
\author{A.~Schwenk}
\email{schwenk@physik.tu-darmstadt.de}
\affiliation{Institut f\"ur Kernphysik, 
Technische Universit\"at Darmstadt, 64289 Darmstadt, Germany}
\affiliation{ExtreMe Matter Institute EMMI, 
GSI Helmholtzzentrum f\"ur Schwerionenforschung GmbH, 64291 Darmstadt,
Germany}
\affiliation{Max-Planck-Institut f\"ur Kernphysik, Saupfercheckweg 1, 
69117 Heidelberg, Germany}

\begin{abstract} 
We present quantum Monte Carlo calculations of few-neutron systems
confined in external potentials based on local chiral interactions at
next-to-next-to-leading order in chiral effective field theory.  The
energy and radial densities for these systems are calculated in
different external Woods-Saxon potentials.  We assume that their
extrapolation to zero external-potential depth provides a quantitative
estimate of three- and four-neutron resonances.  The validity of this
assumption is demonstrated by benchmarking with an exact
diagonalization in the two-body case.  We find that the extrapolated
trineutron resonance, as well as the energy for shallow well depths,
is lower than the tetraneutron resonance energy.  This suggests
that a three-neutron resonance exists below a four-neutron resonance
in nature and is potentially measurable.  To confirm that the relative
ordering of three- and four-neutron resonances is not an artifact of
the external confinement, we test that the odd-even staggering
in the helium isotopic chain is reproduced within this approach.
Finally, we discuss similarities between our results and ultracold
Fermi gases.
\end{abstract}


\maketitle

In recent years, there have been impressive theoretical and
experimental investigations to determine the properties of neutron-rich
nuclei, including isotopic chains of oxygen, calcium, and
others~\cite{Fors13CaEDF,Hebe15ARNPS}. However, understanding
the properties of nuclei beyond the dripline is very challenging and
intriguing. Pure neutron matter has also received much attention, as
it provides a bridge between neutron-rich nuclei, through the symmetry
energy, and neutron
stars~\cite{Tsan12esymm,Tews13N3LO,Hebe13ApJ,Gand15ARNPS}. Therefore,
understanding the interactions between neutrons is an important task.

This question has motivated experimental investigations of few-neutron
systems. In 2002, an experimental claim for a bound tetraneutron
emerged from the detection of neutron clusters from \isotope[14]Be
fragmentation~\cite{marques2002}.  However, this claim has not since
been reproduced, and it seems clear from several increasingly
sophisticated studies~\cite{bertulani2002,timofeyuk2003,pieper2003}
that a tetraneutron system must be unbound.  The possibility of the
existence of a tetraneutron resonance is still an open question.
Recently, a candidate four-neutron resonance has been observed in the
double-charge-exchange reaction
\isotope[4]He(\isotope[8]He,\isotope[8]Be) at an energy of
$(0.83\pm0.65\pm1.25)$~MeV, where the first error is statistical and
the second error is systematic~\cite{kisamori2016}.  Several other
experiments are approved to search for the tetraneutron
resonance~\cite{paschalis_np1406,kisamori_np1512}, including a higher
statistics run of the double-charge-exchange
reaction~\cite{shimoura_np1512}.

On the theoretical side, regarding calculations of a possible
tetraneutron resonance and their sensitivity to nuclear forces, the
situation is inconclusive. Green's function Monte Carlo
calculations~\cite{pieper2003} and no-core-shell-model
calculations~\cite{shirokov2016} suggest that there might be a
tetraneutron resonance with an energy lower than about 2~MeV.  Other
calculations, however, suggest that, in order to have a four-neutron
resonance compatible with the experimental measurements above, the
three-neutron interaction must be strongly modified~\cite{Hiyama:2016},
or even a four-neutron force needs to be invoked~\cite{Lazauskas:2005}.
However, what still remains missing is an \textit{ab initio}
investigation based on two- and three-neutron interactions derived from
chiral effective field theory (EFT). This Letter presents first results
in this direction.

We investigate the properties of two, three, and four neutrons confined in an
external potential.  Our calculations provide evidence that (i) nuclear
Hamiltonians constructed within chiral EFT support a tetraneutron
resonance at an energy of 2.1(2)~MeV compatible with recent
experiments, (ii) because of the extreme diluteness of the system, the
role of three-body (and higher-body) interactions as well as the
effects of details of the regulators in the two-body systems are very
small, (iii) the energy of a three-neutron resonance at 1.1(2)~MeV is
lower than that of four neutrons, and (iv) there are interesting
analogies with systems made of ultracold fermions.  These conclusions
open the possibility for new experimental searches of a trineutron
resonance and that similar systems might be simulated by using
ultracold Fermi gases.

We start our calculation from a many-body Hamiltonian that includes
two- and three-nucleon interactions obtained within the framework of
chiral EFT at next-to-next-to-leading order (N$^2$LO) recently
developed in a local
form~\cite{gezerlis2013,gezerlis2014,lynn2014,tews2016,lynn2016}.
Since the pure neutron system is unbound, we confine the neutrons in
an external trap (called ``neutron drops'').  These systems can be
very accurately solved by starting from microscopic nuclear
Hamiltonians and have been extensively studied with the goal of
improving energy-density functionals in extrapolating to large isospin
asymmetries~\cite{gandolfi2011}.  We model the system starting from
the Hamiltonian
\begin{equation}
H = -\sum_i{\hbar^2\over2m} \laplacian_i
+ \sum_i V_{\rm WS}(r_i) + \sum_{i<j} V_{ij} + \sum_{i<j<k}V_{ijk} \,,
\end{equation} 
where $V_{\rm WS}(r)=-V_0/[1+e^{(r-R_{\rm WS})/a}]$ is a Woods-Saxon
potential with depth $V_0$, radius $R_{\rm WS}$, and diffuseness
$a=0.65$~fm~\cite{BohrMottelson} and $V_{ij}$ and $V_{ijk}$ are two-
and three-body interactions, respectively, constructed at N$^2$LO in
Refs.~\cite{gezerlis2014,tews2016,lynn2016}.  We have checked that our
results are insensitive to the precise value of the diffuseness
parameter $a$.  Changing $a$ by 20\% in either direction changes the
energy by less than 1\% in the two-neutron case.

We use the auxiliary-field diffusion Monte Carlo method
(AFDMC)~\cite{schmidt1999} to project out the ground state from a
variational trial wave function whose form is
\begin{align}
&\langle \vb{R}S|\Psi_V\rangle
\nonumber \\
&=\langle \vb{R}S|
\bigg(\prod_{i<j}f^c(r_{ij})\bigg) 
\bigg(1+\sum_{i<j} F_{ij}
+\sum_{i<j<k} F_{ijk} \bigg)
|\Phi_{JM}\rangle \,,
\end{align}
where $|\vb{R}S\rangle$ represent a collection of sampled $3A$ spatial
coordinates and the $2A$ spinors of the $A$ neutrons with an amplitude
for the $\uparrow$ and $\downarrow$ spin, and $f^c(r_{ij})$, $F_{ij}$,
and $F_{ijk}$ are two- and three-body spin-dependent functions,
respectively, that account for the short-range correlations among
nucleons; see Ref.~\cite{carlson2015} for more details.
$|\Phi\rangle$ is an antisymmetric uncorrelated mean-field part that
describes the correct quantum numbers and asymptotic behavior of the
system.  In our case, it is given by a linear combination of Slater
determinants:
\begin{align} 
\langle\vb{R}S|\Phi_{JM}\rangle 
&= \sum_n k_n\Big[\sum D\{\phi_\alpha(\vb{r}_i,s_i)\}\Big]_{JM} \,,
\nonumber \\
\phi_\alpha(\vb{r}_i,s_i) &= \Phi_{nlj}(r_i)\left[Y_{lm_l}(\vb{\hat{r}}_i)
\xi_{sm_s}(s_i)\right]_{jm_j} \,,
\end{align}
where $[\dots]_{JM}$ means a linear combination of Slater determinants
$D$ coupled with Clebsch-Gordan coefficients to have the quantum
numbers~$JM$.  The radial components $\Phi_{nlj}$ are obtained by
solving the Hartree-Fock equations with the Skyrme force
SKM~\cite{pethick1995}, $Y_{lm_l}$ are spherical harmonics, and
$\xi_{sm_s}$ are spinors in the usual up-down basis.  For each $(JM)$
set of quantum numbers, there are several combinations of
single-particle orbitals.  In particular, we included orbitals in
$1S_{1/2}$, $1P_{3/2}$, $1P_{1/2}$, $1D_{5/2}$, $2S_{1/2}$, and
$1D_{3/2}$.  Since for shallow external potentials the Hartree-Fock
solution is unbound, we tuned the depth of the external trap (imposed on
the orbitals, which is distinct from the external potential) to generate
the orbitals in such a way that they are bound, and then we added an
additional variational parameter to vary their width.  The two- and
three-body correlations as well as the coefficients $k_n$ are obtained
by minimizing the variational energy as described in
Ref.~\cite{sorella2001}.  The ground state of the system is finally
obtained with a projection in imaginary time as
$\Psi(\tau)=\exp[-(H-E_T)\tau]\Psi_V$, where $E_T$ is a parameter that
controls the normalization (the results are independent of the choice
of $E_T$).  In the limit of $\tau\rightarrow\infty$, the lowest energy
state with the symmetry of $\Psi_V$ is found (for more details, see
Ref.~\cite{carlson2015}).
One important point worth emphasizing is that the AFDMC method
does not rely on a basis-set expansion. Therefore, in the
infinite-volume limit, continuum states are automatically included.

\begin{figure}[b]
\begin{center}
\includegraphics[width=1.0\columnwidth]{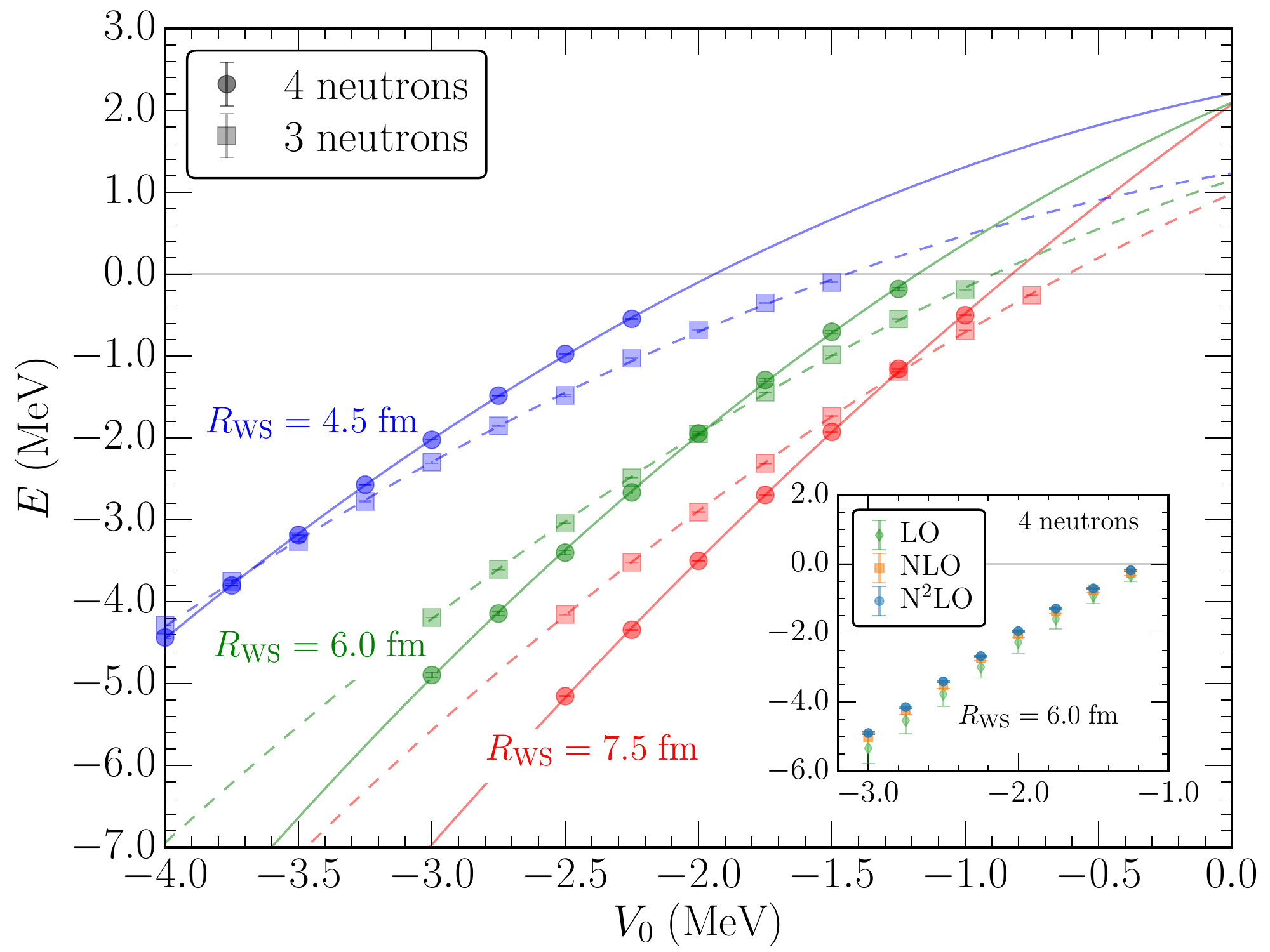}
\end{center}
\caption{The energy of three (squares) and four (circles) neutrons in
external Woods-Saxon potentials for varying radius $R_{\rm WS}$ as a
function of the well depth $V_0$.
The blue (upper) lines correspond to $R_{\rm WS}=5$~fm, the green
(middle) lines to $R_{\rm WS}=6$~fm, and the red (lower) lines to
$R_{\rm WS}=7.5$~fm.
In each case, a quadratic fit to the AFDMC results was obtained and used
to extrapolate to the zero-well-depth limit.
The inset shows calculations of four neutrons at LO
(green diamonds), NLO (orange squares), and N$^2$LO (blue circles) with
uncertainties coming both from the quantum Monte Carlo statistical
uncertainty and from the truncation of the chiral expansion to the order
N$^2$LO (discussed in more detail in the text) for the Woods-Saxon
radius $R_{\rm WS}=6.0$~fm.}
\label{fig:ene}
\end{figure}

We have calculated the energy of three and four neutrons for different
depths $V_0$ and radii $R_{\rm WS}$.  The results are summarized in
Fig.~\ref{fig:ene}, and they have been obtained using the local chiral
potential of Ref.~\cite{lynn2016} with a cutoff of $R_0=1.0$ fm.  The
plot shows the energy as a function of $V_0$ for three~(squares) and
four~(circles) neutrons.  The blue~(upper curves for various neutron
numbers), green~(middle~curves), and red~(lower~curves) are the
results obtained for different radii $R_{\rm WS}$ as indicated in the
figure.  The lines are quadratic fits to the energies of
four~(solid~lines) and three~(dashed~lines) neutrons.  The
extrapolations to $V_0\rightarrow 0$ obtained for the different values
of $R_{\rm WS}$ converge to the same point, indicating that the
results at zero well depth are independent of the geometry of the
external potential (provided that it goes to zero at large distances
and its range is larger than the nucleon-nucleon effective range).
Since we are simulating a system that is naturally unbound,
we enforce the center of mass to have no motion in order to calculate
internal energies only, as is commonly done in quantum Monte Carlo
calculations for nuclei; i.e., given the translationally invariant
Hamiltonian, the Monte Carlo evolution is performed so that the center
of mass of the system does not move.

In order to establish the role of the cutoff $R_0$ in the
nucleon-nucleon interaction and that of the three-body forces, we have
repeated the calculation using $R_0=1.2$~fm and turning off the
three-neutron interaction.  The results are indistinguishable from the
cases shown in Fig.~\ref{fig:ene}, within statistical errors (which are
smaller than the points).  Given the density of the system, this is not
totally unexpected, as we discuss below.  Another source of systematic
uncertainty comes from the truncation of the chiral expansion at
N$^2$LO.  To estimate this uncertainty, we have considered the case of
four neutrons in the Woods-Saxon well with $R_\text{WS}=6.0$~fm and
repeated our calculations at leading order (LO) and next-to-leading
order (NLO).  Following Ref.~\cite{epelbaum2015}, we estimate the
uncertainty coming from the truncation of the chiral expansion at
N$^2$LO.  We add these in quadrature to the quantum Monte Carlo
statistical uncertainties.  These are displayed as the error bars in
Fig.~\ref{fig:ene} for the case with $R_\text{WS}=6.0$~fm.  They are
still smaller than the points and, within the uncertainties we have
quoted, do not affect the extrapolated energy of the four-neutron
system.  The inset also shows the LO, NLO, and N$^2$LO results with
uncertainties as described above.  One can see that, especially near the
limit where the system becomes unbound, the results are not very
sensitive to the chiral order. The fits in Fig.~\ref{fig:ene} give an
energy per particle of 0.37(7)~MeV for three neutrons and 0.53(5)~MeV
for four neutrons.  This suggests that there could be a trineutron
resonance in nature at a lower energy than the four-neutron resonance.
We have also considered the extrapolation from a different
approach. We have multiplied the N$^2$LO interaction by an overall scale
factor $\alpha$ and tuned $\alpha$ until the four neutrons were bound as
in Ref.~\cite{fossez2016}. We find a scale factor of $\alpha\sim1.3$ is
sufficient to bind the four neutrons.  We have varied $\alpha$ and
performed an extrapolation similar to what is shown in
Fig.~\ref{fig:ene} and found an energy for the unbound system at
$\alpha=1$ of $E=2.0(1.0)$~MeV, which is consistent with our results
coming from the trapped four neutrons.

\begin{figure}[t]
\begin{center}
\includegraphics[width=1.0\columnwidth]{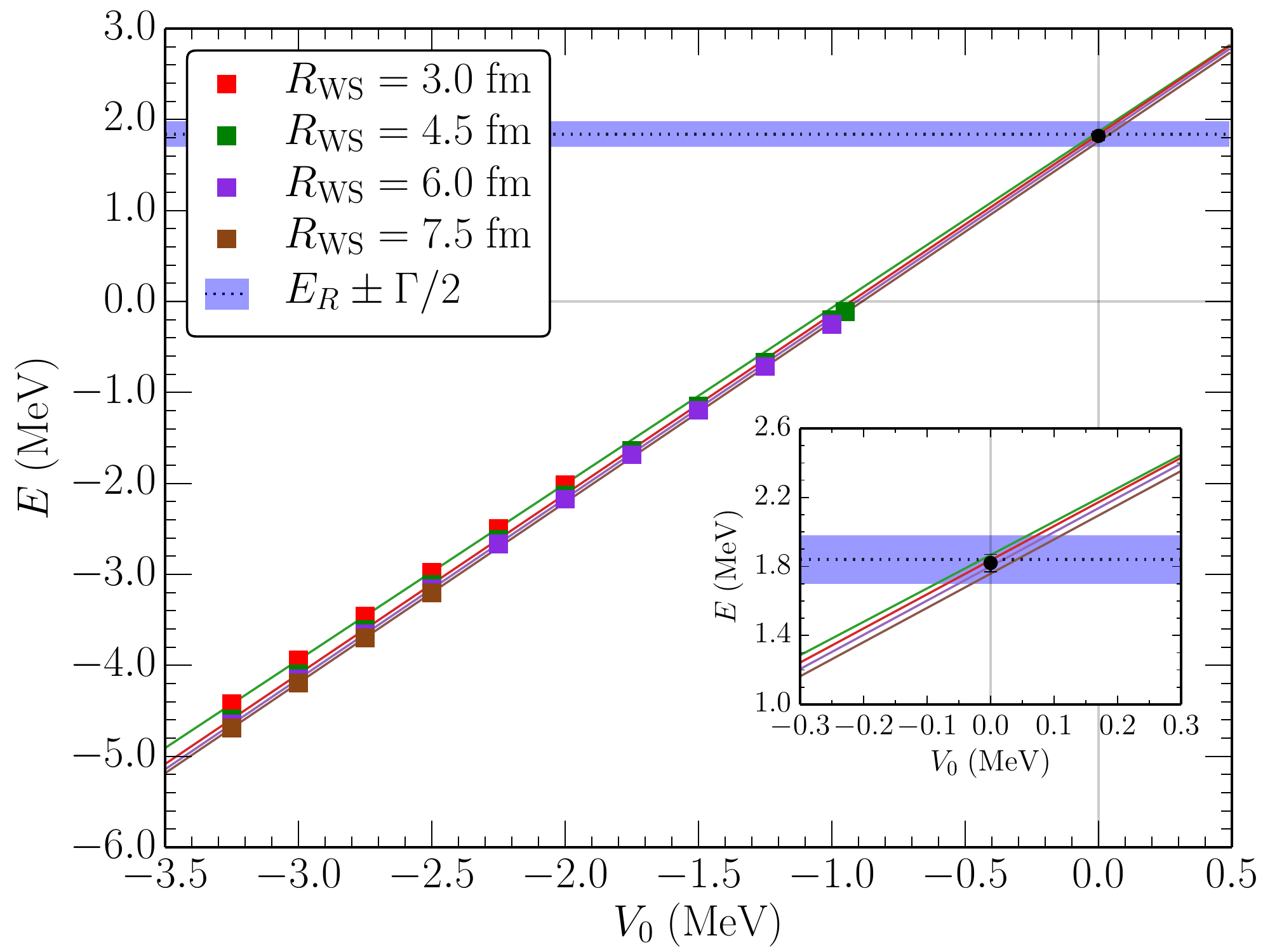}
\end{center}
\caption{Energy of two neutrons trapped in various Woods-Saxon wells
interacting via a simple model potential~Eq.(\ref{eq:2gauss}) designed
to give a low-lying resonance.  Also shown are the linear
extrapolations to zero well depth and the resonance energy $E_R$ and
width $\Gamma$ extracted from the continuum.  The black point at
$V_0=0$~MeV is the average and standard deviation of the
extrapolations evaluated at zero well depth.}
\label{fig:nntoy_extrap}
\end{figure}

Our results rely on the assumption that the extrapolation of the
energy to the zero-depth external potential may be interpreted as a
resonance energy, as suggested in Ref.~\cite{pieper2003}.  To provide
support for this interpretation, we have designed a simple $S$-wave
potential consisting of two Gaussians:
\begin{equation}
\label{eq:2gauss}
V(r)=V_1 \, \eu{-\left(\frac{r}{R_1}\right)^2}
+V_2 \, \eu{-\left(\frac{r-r_2}{R_2}\right)^2} \,,
\end{equation}
with parameters $V_1=-1000$~MeV, $V_2=865$~MeV, $R_1=0.4981$~fm,
$R_2=0.2877$~fm, and $r_2=0.9972$~fm, such that we have an attractive
well at the origin and a repulsive barrier at $\sim1.0$~fm.  This
potential gives a resonance at $E_R=1.84$~MeV with a width of
$\Gamma=0.282$~MeV. In Fig.~\ref{fig:nntoy_extrap}, we have
diagonalized the two-body Hamiltonian with this simple $S$-wave
potential plus Woods-Saxon wells of various widths and depths.
Extrapolating the bound-state energies to zero well depth as in the
realistic case, we have found an energy intercept $E_R=1.83(5)$~MeV.
Similarly, we have constructed a two-body interaction that does not
have any resonance (a purely attractive Gaussian) and found that the
Woods-Saxon depth required to bind the system is unnaturally large
and that the extrapolations for individual widths do not converge to
the same energy at zero well depth.  In addition, we have calculated
the energy of two neutrons interacting via the chiral N$^2$LO
interactions in a Woods-Saxon well and found an extrapolation
compatible with the virtual state energy of $\sim0.1$~MeV.  These
exact calculations therefore provide evidence that our extrapolation
method can provide meaningful resonance energies.

\begin{figure}[t]
\begin{center}
\includegraphics[width=1.0\columnwidth]{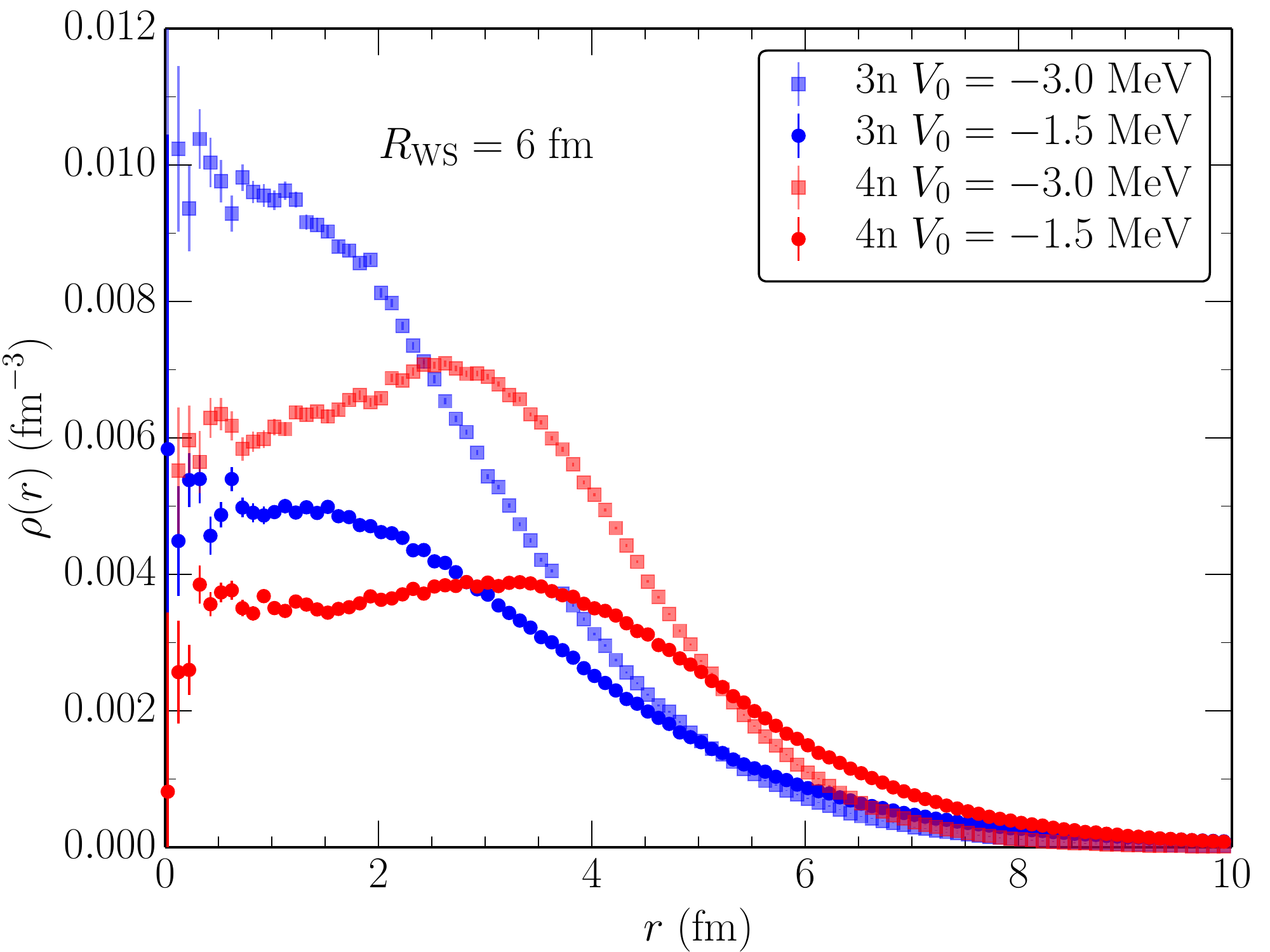}
\end{center}
\caption{One-body densities for three (blue) and four (red) neutrons
in two different Woods-Saxon wells with depths 3~MeV (squares) and
1.5~MeV (circles) with a fixed $R_{\rm WS}=6.0$~fm.}
\label{fig:dens}
\end{figure}
\begin{figure}[b]
\begin{center}
\includegraphics[width=1.0\columnwidth]{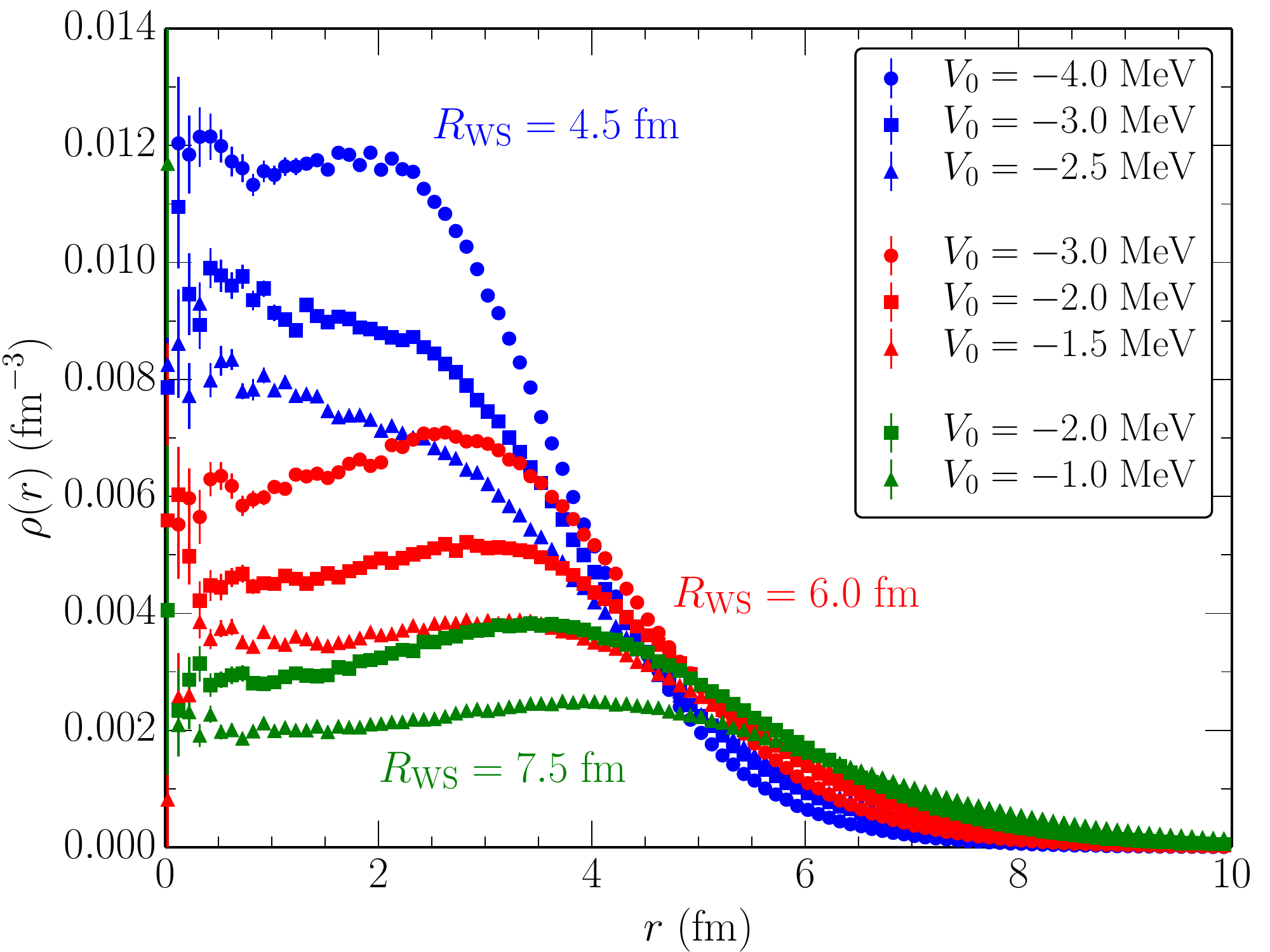}
\end{center}
\caption{One-body densities for four neutrons in Woods-Saxon wells
with various depths and widths.}
\label{fig:densn4}
\end{figure}

\begin{figure}[b]
\begin{center}
\includegraphics[width=1.0\columnwidth]{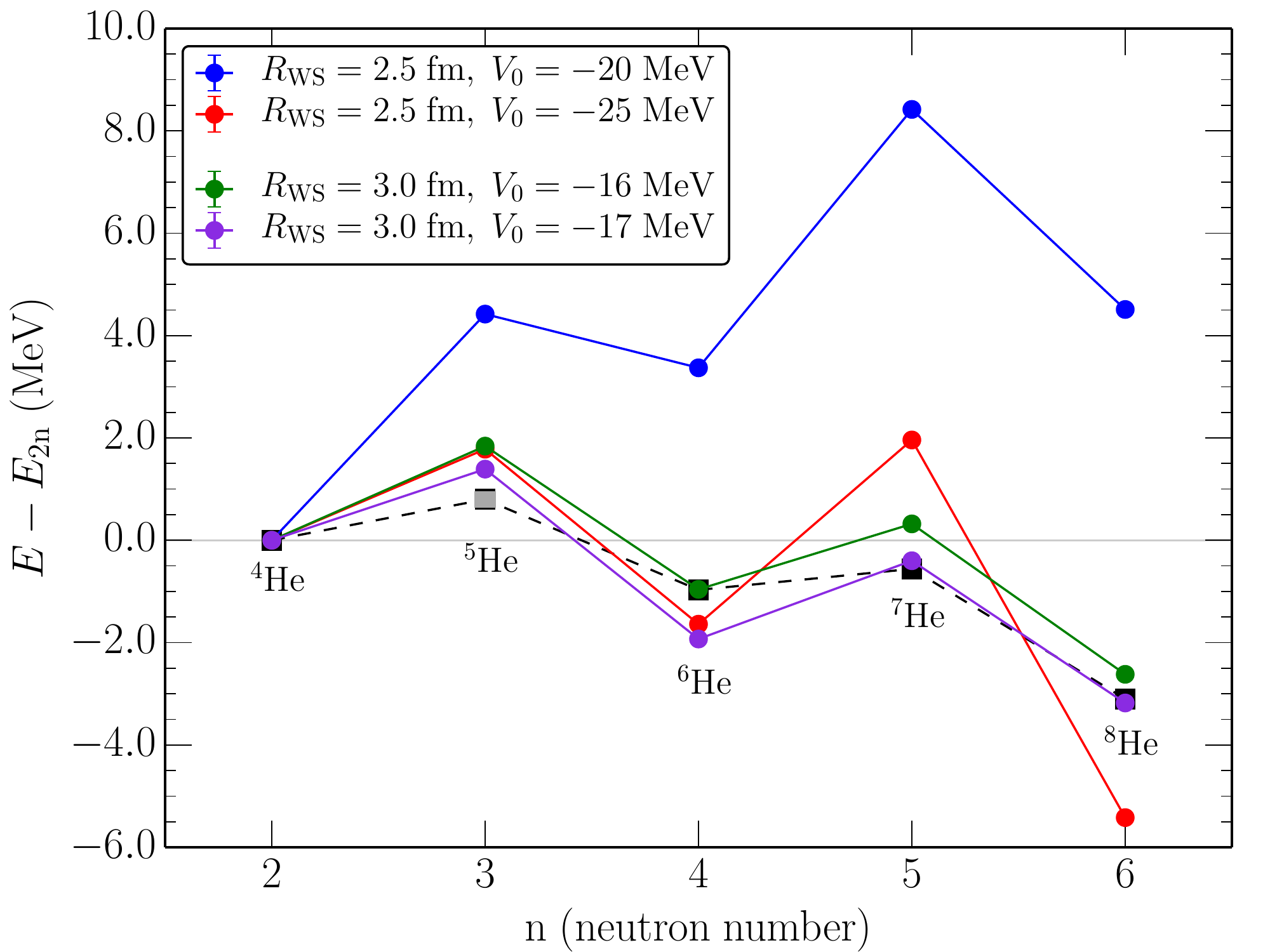}
\end{center}
\caption{Energy of two to six neutrons trapped in various Woods-Saxon
wells (circles). The wells are designed to approximately reproduce
the binding pattern of the helium chain.  For each well, the two-neutron
energy is taken as the reference point to which the other energies for
that well are compared.  The black squares are the experimental values
compared to the \isotope[4]{He} energy.
For \isotope[5]{He}, we take the value of the $P_{\sfrac{3}{2}}$
resonance, the width of which is shown in gray.
\label{fig:hechain}}
\end{figure}

We have also computed the density distribution of neutrons in the
trap.  In Fig.~\ref{fig:dens}, we show the neutron distribution inside
the trap for three and four neutrons in different Woods-Saxon wells
with $R_{\rm WS}=6$~fm, normalized such that their integral is equal
to the number of neutrons.  As can been seen, the density of the
systems never exceeds the value of $\sim0.01$~fm$^{-3}$, suggesting
that the system is very dilute.  In the case of infinite neutron
matter~\cite{gezerlis2014,tews2016,lynn2016}, at such low densities
the energy per neutron is totally dominated by the $S$-wave part of
the neutron-neutron interaction, and the results are almost
independent of the two-body cutoff $R_0$ and the three-neutron
interaction.  However, it is interesting to note that in the same well
the three-neutron system is always denser near the center than the
four-neutron system, and the latter shows a distribution with a peak
around 3~fm, suggesting that the system is arranged on a ``shell.''
Notably, this difference in shape between the three- and
four-neutron systems persists as the geometry of the trap is changed.
One possible interpretation is that in the case of three neutrons one
pair (up-down) of neutrons is sitting in the center of the trap, and
one extra neutron is orbiting around in a $P$ state.  In the case of
four neutrons, instead, the two pairs are orbiting around the center,
making the system less dense in the center.  It would be very
interesting to measure these properties by tracking the position of
the neutrons.  The density of four neutrons in Woods-Saxon wells with
different $V_0$ and $R_{\rm WS}$ is shown in Fig.~\ref{fig:densn4}.
Also in this case we can verify that the system is very dilute.

Finally, we have performed additional calculations of two to six
neutrons in different wells adapted to qualitatively mimic the helium
isotopes.  In this model, we replace the two protons with a
Woods-Saxon potential and calculate the energy of neutrons in such a
well, interacting with the N$^2$LO interaction.  This model has been
successfully applied to describe the oxygen isotopic
chain~\cite{Chang:2004}.  In Fig.~\ref{fig:hechain}, we show the
energy of the helium isotopic chain as obtained from this simplified
model.  The results are normalized to the \isotope[4]{He} energy,
which corresponds to the energy of only two neutrons in the
Woods-Saxon well.  Again, we keep the center of mass of the system
fixed.  Considering different Woods-Saxon potentials, we find in this
case the expected odd-even pairing effects; i.e., the systems with odd
numbers of neutrons always have higher energies than the neighboring
systems with an even number of neutrons. In this case, $V_0$ is
strongly attractive, and compared to Fig.~\ref{fig:ene} the
ordering of three versus four neutron energies is reversed. For the
helium isotopes, we attribute this to the additional pairing
attraction generated from interacting with the $^4$He core. The
ordering, with a lower trineutron energy, changes in the region of small
$V_0$ where densities are much lower than for the helium isotopes.

Our results can be interpreted from the viewpoint of ultracold atom
experiments. We observe that the extrapolated resonance energies of
three- and four-neutron states in Fig.~\ref{fig:ene} scale with the
number of pairs, which is $N(N-1)/2$.  This behavior can be
qualitatively understood by considering the diluteness of the system.
For a large particle number $N$, the scaling with the number of pairs is
consistent with the scaling of the mean-field (MF) interaction energy
of a dilute gas of spin-1/2 fermions~\cite{FW:1971}:
\begin{equation}
E_\text{MF} = \frac{\pi a}{m} \frac{N^2}{V} \,,
\end{equation}
which scales as $N^2$. Here, $a$ is the two-body scattering length and
$V$ is the volume. Quantum degenerate Fermi gases can also be
engineered in experiments with ultracold atoms~\cite{deMarco:1999}.
The mean-field energy of a two-component Fermi gas in a harmonic trap
was measured for both signs of the scattering length using
radio-frequency spectroscopy~\cite{Regal:2003}.  This suggests that
few-neutron resonances and the transition from few- to many-body
physics could be simulated in experiments with ultracold atoms.
Similar experiments have already been carried out for
quasi-one-dimensional systems with an impurity, where it was found that
systems with $N\geq4$ majority atoms already develop a Fermi
sea~\cite{Zuern:2013}.  Moreover, experiments with ultracold atoms
could be used to investigate whether the properties of the density
distributions in Figs.~\ref{fig:dens} and~\ref{fig:densn4} are
governed by universal large-scattering length physics or details of
nuclear forces.

In this Letter, we have simulated two, three, and four neutrons in
external potentials and extrapolated to the zero-well-depth limit.
These extrapolations are independent of the trap geometry, since
different Woods-Saxon widths converge to the same energy at zero well
depth.  We found a tetraneutron resonance energy in agreement with
recent measurements.  Taken together with the results from the simple
$S$-wave potential and the results mimicking the helium isotopic
chain, our results suggest that a trineutron resonance may be lower in
energy than a four-neutron resonance and therefore possibly
experimentally observable.  We also conclude that the effects of
three-neutron interactions are very small in these systems due to
their diluteness.  In addition, the diluteness of these
systems offers the exciting possibility to shed more light on the
properties of few-neutron systems with experiments with ultracold atomic
Fermi gases.

\acknowledgments{We thank T.~Aumann, J.~A.~Carlson, B.~Gibson,
S.~Pieper, A.~Rusetsky, K.~E.~Schmidt and R.~B.~Wiringa for useful
discussions. This work was supported by the US Department of Energy
under Contract No.~DE-AC52-06NA25396, the NUCLEI SciDAC project, the ERC
Grant No.~307986 STRONGINT, and the Deutsche Forschungsgemeinschaft
through Grant SFB~1245. The computations were performed at Los Alamos
Open Supercomputing, the J\"ulich Supercomputing Center, NERSC, which is
supported by the US Department of Energy under Contract
No.~DE-AC02-05CH11231, and on the Lichtenberg high performance
computer of the TU Darmstadt.}

\bibliography{neutres11}

\begin{thebibliography}{35}%
\makeatletter
\providecommand \@ifxundefined [1]{%
 \@ifx{#1\undefined}
}%
\providecommand \@ifnum [1]{%
 \ifnum #1\expandafter \@firstoftwo
 \else \expandafter \@secondoftwo
 \fi
}%
\providecommand \@ifx [1]{%
 \ifx #1\expandafter \@firstoftwo
 \else \expandafter \@secondoftwo
 \fi
}%
\providecommand \natexlab [1]{#1}%
\providecommand \enquote  [1]{``#1''}%
\providecommand \bibnamefont  [1]{#1}%
\providecommand \bibfnamefont [1]{#1}%
\providecommand \citenamefont [1]{#1}%
\providecommand \href@noop [0]{\@secondoftwo}%
\providecommand \href [0]{\begingroup \@sanitize@url \@href}%
\providecommand \@href[1]{\@@startlink{#1}\@@href}%
\providecommand \@@href[1]{\endgroup#1\@@endlink}%
\providecommand \@sanitize@url [0]{\catcode `\\12\catcode `\$12\catcode
  `\&12\catcode `\#12\catcode `\^12\catcode `\_12\catcode `\%12\relax}%
\providecommand \@@startlink[1]{}%
\providecommand \@@endlink[0]{}%
\providecommand \url  [0]{\begingroup\@sanitize@url \@url }%
\providecommand \@url [1]{\endgroup\@href {#1}{\urlprefix }}%
\providecommand \urlprefix  [0]{URL }%
\providecommand \Eprint [0]{\href }%
\providecommand \doibase [0]{http://dx.doi.org/}%
\providecommand \selectlanguage [0]{\@gobble}%
\providecommand \bibinfo  [0]{\@secondoftwo}%
\providecommand \bibfield  [0]{\@secondoftwo}%
\providecommand \translation [1]{[#1]}%
\providecommand \BibitemOpen [0]{}%
\providecommand \bibitemStop [0]{}%
\providecommand \bibitemNoStop [0]{.\EOS\space}%
\providecommand \EOS [0]{\spacefactor3000\relax}%
\providecommand \BibitemShut  [1]{\csname bibitem#1\endcsname}%
\let\auto@bib@innerbib\@empty
\bibitem [{\citenamefont {Forss{\'e}n}\ \emph {et~al.}(2013)\citenamefont
  {Forss{\'e}n}, \citenamefont {Hagen}, \citenamefont {Hjorth-Jensen},
  \citenamefont {Nazarewicz},\ and\ \citenamefont {Rotureau}}]{Fors13CaEDF}%
  \BibitemOpen
  \bibfield  {author} {\bibinfo {author} {\bibfnamefont {C.}~\bibnamefont
  {Forss{\'e}n}}, \bibinfo {author} {\bibfnamefont {G.}~\bibnamefont {Hagen}},
  \bibinfo {author} {\bibfnamefont {M.}~\bibnamefont {Hjorth-Jensen}}, \bibinfo
  {author} {\bibfnamefont {W.}~\bibnamefont {Nazarewicz}}, \ and\ \bibinfo
  {author} {\bibfnamefont {J.}~\bibnamefont {Rotureau}},\ }\href@noop {}
  {\bibfield  {journal} {\bibinfo  {journal} {Phys. Scr. T}\ }\textbf {\bibinfo
  {volume} {152}},\ \bibinfo {pages} {014022} (\bibinfo {year}
  {2013})}\BibitemShut {NoStop}%
\bibitem [{\citenamefont {Hebeler}\ \emph {et~al.}(2015)\citenamefont
  {Hebeler}, \citenamefont {Holt}, \citenamefont {Men\'endez},\ and\
  \citenamefont {Schwenk}}]{Hebe15ARNPS}%
  \BibitemOpen
  \bibfield  {author} {\bibinfo {author} {\bibfnamefont {K.}~\bibnamefont
  {Hebeler}}, \bibinfo {author} {\bibfnamefont {J.~D.}\ \bibnamefont {Holt}},
  \bibinfo {author} {\bibfnamefont {J.}~\bibnamefont {Men\'endez}}, \ and\
  \bibinfo {author} {\bibfnamefont {A.}~\bibnamefont {Schwenk}},\ }\href@noop
  {} {\bibfield  {journal} {\bibinfo  {journal} {Annu. Rev. Nucl. Part. Sci.}\
  }\textbf {\bibinfo {volume} {65}},\ \bibinfo {pages} {457} (\bibinfo {year}
  {2015})}\BibitemShut {NoStop}%
\bibitem [{\citenamefont {Tsang}\ \emph {et~al.}(2012)\citenamefont {Tsang}
  \emph {et~al.}}]{Tsan12esymm}%
  \BibitemOpen
  \bibfield  {author} {\bibinfo {author} {\bibfnamefont {M.~B.}\ \bibnamefont
  {Tsang}} \emph {et~al.},\ }\href@noop {} {\bibfield  {journal} {\bibinfo
  {journal} {Phys. Rev. C}\ }\textbf {\bibinfo {volume} {86}},\ \bibinfo
  {pages} {015803} (\bibinfo {year} {2012})}\BibitemShut {NoStop}%
\bibitem [{\citenamefont {Tews}\ \emph {et~al.}(2013)\citenamefont {Tews},
  \citenamefont {Kr{\"u}ger}, \citenamefont {Hebeler},\ and\ \citenamefont
  {Schwenk}}]{Tews13N3LO}%
  \BibitemOpen
  \bibfield  {author} {\bibinfo {author} {\bibfnamefont {I.}~\bibnamefont
  {Tews}}, \bibinfo {author} {\bibfnamefont {T.}~\bibnamefont {Kr{\"u}ger}},
  \bibinfo {author} {\bibfnamefont {K.}~\bibnamefont {Hebeler}}, \ and\
  \bibinfo {author} {\bibfnamefont {A.}~\bibnamefont {Schwenk}},\ }\href@noop
  {} {\bibfield  {journal} {\bibinfo  {journal} {Phys. Rev. Lett.}\ }\textbf
  {\bibinfo {volume} {110}},\ \bibinfo {pages} {032504} (\bibinfo {year}
  {2013})}\BibitemShut {NoStop}%
\bibitem [{\citenamefont {Hebeler}\ \emph {et~al.}(2013)\citenamefont
  {Hebeler}, \citenamefont {Lattimer}, \citenamefont {Pethick},\ and\
  \citenamefont {Schwenk}}]{Hebe13ApJ}%
  \BibitemOpen
  \bibfield  {author} {\bibinfo {author} {\bibfnamefont {K.}~\bibnamefont
  {Hebeler}}, \bibinfo {author} {\bibfnamefont {J.~M.}\ \bibnamefont
  {Lattimer}}, \bibinfo {author} {\bibfnamefont {C.~J.}\ \bibnamefont
  {Pethick}}, \ and\ \bibinfo {author} {\bibfnamefont {A.}~\bibnamefont
  {Schwenk}},\ }\href@noop {} {\bibfield  {journal} {\bibinfo  {journal}
  {Astrophys. J.}\ }\textbf {\bibinfo {volume} {773}},\ \bibinfo {pages} {11}
  (\bibinfo {year} {2013})}\BibitemShut {NoStop}%
\bibitem [{\citenamefont {Gandolfi}\ \emph {et~al.}(2015)\citenamefont
  {Gandolfi}, \citenamefont {Gezerlis},\ and\ \citenamefont
  {Carlson}}]{Gand15ARNPS}%
  \BibitemOpen
  \bibfield  {author} {\bibinfo {author} {\bibfnamefont {S.}~\bibnamefont
  {Gandolfi}}, \bibinfo {author} {\bibfnamefont {A.}~\bibnamefont {Gezerlis}},
  \ and\ \bibinfo {author} {\bibfnamefont {J.~A.}\ \bibnamefont {Carlson}},\
  }\href@noop {} {\bibfield  {journal} {\bibinfo  {journal} {Annu. Rev. Nucl.
  Part. Sci.}\ }\textbf {\bibinfo {volume} {65}},\ \bibinfo {pages} {303}
  (\bibinfo {year} {2015})}\BibitemShut {NoStop}%
\bibitem [{\citenamefont {Marqu\'es}\ \emph {et~al.}(2002)\citenamefont
  {Marqu\'es} \emph {et~al.}}]{marques2002}%
  \BibitemOpen
  \bibfield  {author} {\bibinfo {author} {\bibfnamefont {F.~M.}\ \bibnamefont
  {Marqu\'es}} \emph {et~al.},\ }\href {\doibase 10.1103/PhysRevC.65.044006}
  {\bibfield  {journal} {\bibinfo  {journal} {Phys. Rev. C}\ }\textbf {\bibinfo
  {volume} {65}},\ \bibinfo {pages} {044006} (\bibinfo {year}
  {2002})}\BibitemShut {NoStop}%
\bibitem [{\citenamefont {Bertulani}\ and\ \citenamefont
  {Zelevinsky}(2003)}]{bertulani2002}%
  \BibitemOpen
  \bibfield  {author} {\bibinfo {author} {\bibfnamefont {C.~A.}\ \bibnamefont
  {Bertulani}}\ and\ \bibinfo {author} {\bibfnamefont {V.}~\bibnamefont
  {Zelevinsky}},\ }\href {\doibase 10.1088/0954-3899/29/10/309} {\bibfield
  {journal} {\bibinfo  {journal} {J. Phys. G}\ }\textbf {\bibinfo {volume}
  {29}},\ \bibinfo {pages} {2431} (\bibinfo {year} {2003})}\BibitemShut
  {NoStop}%
\bibitem [{\citenamefont {Timofeyuk}(2003)}]{timofeyuk2003}%
  \BibitemOpen
  \bibfield  {author} {\bibinfo {author} {\bibfnamefont {N.~K.}\ \bibnamefont
  {Timofeyuk}},\ }\href {\doibase 10.1088/0954-3899/29/2/102} {\bibfield
  {journal} {\bibinfo  {journal} {J. Phys. G}\ }\textbf {\bibinfo {volume}
  {29}},\ \bibinfo {pages} {L9} (\bibinfo {year} {2003})}\BibitemShut {NoStop}%
\bibitem [{\citenamefont {Pieper}(2003)}]{pieper2003}%
  \BibitemOpen
  \bibfield  {author} {\bibinfo {author} {\bibfnamefont {S.~C.}\ \bibnamefont
  {Pieper}},\ }\href {\doibase 10.1103/PhysRevLett.90.252501} {\bibfield
  {journal} {\bibinfo  {journal} {Phys. Rev. Lett.}\ }\textbf {\bibinfo
  {volume} {90}},\ \bibinfo {pages} {252501} (\bibinfo {year}
  {2003})}\BibitemShut {NoStop}%
\bibitem [{\citenamefont {Kisamori}\ \emph {et~al.}(2016)\citenamefont
  {Kisamori} \emph {et~al.}}]{kisamori2016}%
  \BibitemOpen
  \bibfield  {author} {\bibinfo {author} {\bibfnamefont {K.}~\bibnamefont
  {Kisamori}} \emph {et~al.},\ }\href {\doibase 10.1103/PhysRevLett.116.052501}
  {\bibfield  {journal} {\bibinfo  {journal} {Phys. Rev. Lett.}\ }\textbf
  {\bibinfo {volume} {116}},\ \bibinfo {pages} {052501} (\bibinfo {year}
  {2016})}\BibitemShut {NoStop}%
\bibitem [{\citenamefont {Paschalis}\ \emph {et~al.}()\citenamefont {Paschalis}
  \emph {et~al.}}]{paschalis_np1406}%
  \BibitemOpen
  \bibfield  {author} {\bibinfo {author} {\bibfnamefont {S.}~\bibnamefont
  {Paschalis}} \emph {et~al.},\ }\href@noop {} {\bibinfo  {journal} {Report No.
  NP1406-SAMURAI19}\ }\BibitemShut {NoStop}%
\bibitem [{\citenamefont {Kisamori}\ \emph {et~al.}()\citenamefont {Kisamori}
  \emph {et~al.}}]{kisamori_np1512}%
  \BibitemOpen
\bibfield  {journal} {  }\bibfield  {author} {\bibinfo {author} {\bibfnamefont
  {K.}~\bibnamefont {Kisamori}} \emph {et~al.},\ }\href@noop {} {\bibinfo
  {journal} {Report No. NP1512-SAMURAI34}\ }\BibitemShut {NoStop}%
\bibitem [{\citenamefont {Shimoura}\ \emph {et~al.}()\citenamefont {Shimoura}
  \emph {et~al.}}]{shimoura_np1512}%
  \BibitemOpen
\bibfield  {journal} {  }\bibfield  {author} {\bibinfo {author} {\bibfnamefont
  {S.}~\bibnamefont {Shimoura}} \emph {et~al.},\ }\href@noop {} {\bibinfo
  {journal} {Report No. NP1512-SHARAQ10}\ }\BibitemShut {NoStop}%
\bibitem [{\citenamefont {Shirokov}\ \emph {et~al.}(2016)\citenamefont
  {Shirokov}, \citenamefont {Papadimitriou}, \citenamefont {Mazur},
  \citenamefont {Mazur}, \citenamefont {Roth},\ and\ \citenamefont
  {Vary}}]{shirokov2016}%
  \BibitemOpen
\bibfield  {journal} {  }\bibfield  {author} {\bibinfo {author} {\bibfnamefont
  {A.~M.}\ \bibnamefont {Shirokov}}, \bibinfo {author} {\bibfnamefont
  {G.}~\bibnamefont {Papadimitriou}}, \bibinfo {author} {\bibfnamefont {A.~I.}\
  \bibnamefont {Mazur}}, \bibinfo {author} {\bibfnamefont {I.~A.}\ \bibnamefont
  {Mazur}}, \bibinfo {author} {\bibfnamefont {R.}~\bibnamefont {Roth}}, \ and\
  \bibinfo {author} {\bibfnamefont {J.~P.}\ \bibnamefont {Vary}},\ }\href
  {\doibase 10.1103/PhysRevLett.117.182502} {\bibfield  {journal} {\bibinfo
  {journal} {Phys. Rev. Lett.}\ }\textbf {\bibinfo {volume} {117}},\ \bibinfo
  {pages} {182502} (\bibinfo {year} {2016})}\BibitemShut {NoStop}%
\bibitem [{\citenamefont {Hiyama}\ \emph {et~al.}(2016)\citenamefont {Hiyama},
  \citenamefont {Lazauskas}, \citenamefont {Carbonell},\ and\ \citenamefont
  {Kamimura}}]{Hiyama:2016}%
  \BibitemOpen
  \bibfield  {author} {\bibinfo {author} {\bibfnamefont {E.}~\bibnamefont
  {Hiyama}}, \bibinfo {author} {\bibfnamefont {R.}~\bibnamefont {Lazauskas}},
  \bibinfo {author} {\bibfnamefont {J.}~\bibnamefont {Carbonell}}, \ and\
  \bibinfo {author} {\bibfnamefont {M.}~\bibnamefont {Kamimura}},\ }\href
  {\doibase 10.1103/PhysRevC.93.044004} {\bibfield  {journal} {\bibinfo
  {journal} {Phys. Rev. C}\ }\textbf {\bibinfo {volume} {93}},\ \bibinfo
  {pages} {044004} (\bibinfo {year} {2016})}\BibitemShut {NoStop}%
\bibitem [{\citenamefont {Lazauskas}\ and\ \citenamefont
  {Carbonell}(2005)}]{Lazauskas:2005}%
  \BibitemOpen
  \bibfield  {author} {\bibinfo {author} {\bibfnamefont {R.}~\bibnamefont
  {Lazauskas}}\ and\ \bibinfo {author} {\bibfnamefont {J.}~\bibnamefont
  {Carbonell}},\ }\href {\doibase 10.1103/PhysRevC.72.034003} {\bibfield
  {journal} {\bibinfo  {journal} {Phys. Rev. C}\ }\textbf {\bibinfo {volume}
  {72}},\ \bibinfo {pages} {034003} (\bibinfo {year} {2005})}\BibitemShut
  {NoStop}%
\bibitem [{\citenamefont {Gezerlis}\ \emph {et~al.}(2013)\citenamefont
  {Gezerlis}, \citenamefont {Tews}, \citenamefont {Epelbaum}, \citenamefont
  {Gandolfi}, \citenamefont {Hebeler}, \citenamefont {Nogga},\ and\
  \citenamefont {Schwenk}}]{gezerlis2013}%
  \BibitemOpen
  \bibfield  {author} {\bibinfo {author} {\bibfnamefont {A.}~\bibnamefont
  {Gezerlis}}, \bibinfo {author} {\bibfnamefont {I.}~\bibnamefont {Tews}},
  \bibinfo {author} {\bibfnamefont {E.}~\bibnamefont {Epelbaum}}, \bibinfo
  {author} {\bibfnamefont {S.}~\bibnamefont {Gandolfi}}, \bibinfo {author}
  {\bibfnamefont {K.}~\bibnamefont {Hebeler}}, \bibinfo {author} {\bibfnamefont
  {A.}~\bibnamefont {Nogga}}, \ and\ \bibinfo {author} {\bibfnamefont
  {A.}~\bibnamefont {Schwenk}},\ }\href {\doibase
  10.1103/PhysRevLett.111.032501} {\bibfield  {journal} {\bibinfo  {journal}
  {Phys. Rev. Lett.}\ }\textbf {\bibinfo {volume} {111}},\ \bibinfo {pages}
  {032501} (\bibinfo {year} {2013})}\BibitemShut {NoStop}%
\bibitem [{\citenamefont {Gezerlis}\ \emph {et~al.}(2014)\citenamefont
  {Gezerlis}, \citenamefont {Tews}, \citenamefont {Epelbaum}, \citenamefont
  {Freunek}, \citenamefont {Gandolfi}, \citenamefont {Hebeler}, \citenamefont
  {Nogga},\ and\ \citenamefont {Schwenk}}]{gezerlis2014}%
  \BibitemOpen
  \bibfield  {author} {\bibinfo {author} {\bibfnamefont {A.}~\bibnamefont
  {Gezerlis}}, \bibinfo {author} {\bibfnamefont {I.}~\bibnamefont {Tews}},
  \bibinfo {author} {\bibfnamefont {E.}~\bibnamefont {Epelbaum}}, \bibinfo
  {author} {\bibfnamefont {M.}~\bibnamefont {Freunek}}, \bibinfo {author}
  {\bibfnamefont {S.}~\bibnamefont {Gandolfi}}, \bibinfo {author}
  {\bibfnamefont {K.}~\bibnamefont {Hebeler}}, \bibinfo {author} {\bibfnamefont
  {A.}~\bibnamefont {Nogga}}, \ and\ \bibinfo {author} {\bibfnamefont
  {A.}~\bibnamefont {Schwenk}},\ }\href {\doibase 10.1103/PhysRevC.90.054323}
  {\bibfield  {journal} {\bibinfo  {journal} {Phys. Rev. C}\ }\textbf {\bibinfo
  {volume} {90}},\ \bibinfo {pages} {054323} (\bibinfo {year}
  {2014})}\BibitemShut {NoStop}%
\bibitem [{\citenamefont {Lynn}\ \emph {et~al.}(2014)\citenamefont {Lynn},
  \citenamefont {Carlson}, \citenamefont {Epelbaum}, \citenamefont {Gandolfi},
  \citenamefont {Gezerlis},\ and\ \citenamefont {Schwenk}}]{lynn2014}%
  \BibitemOpen
  \bibfield  {author} {\bibinfo {author} {\bibfnamefont {J.~E.}\ \bibnamefont
  {Lynn}}, \bibinfo {author} {\bibfnamefont {J.}~\bibnamefont {Carlson}},
  \bibinfo {author} {\bibfnamefont {E.}~\bibnamefont {Epelbaum}}, \bibinfo
  {author} {\bibfnamefont {S.}~\bibnamefont {Gandolfi}}, \bibinfo {author}
  {\bibfnamefont {A.}~\bibnamefont {Gezerlis}}, \ and\ \bibinfo {author}
  {\bibfnamefont {A.}~\bibnamefont {Schwenk}},\ }\href {\doibase
  10.1103/PhysRevLett.113.192501} {\bibfield  {journal} {\bibinfo  {journal}
  {Phys. Rev. Lett.}\ }\textbf {\bibinfo {volume} {113}},\ \bibinfo {pages}
  {192501} (\bibinfo {year} {2014})}\BibitemShut {NoStop}%
\bibitem [{\citenamefont {Tews}\ \emph {et~al.}(2016)\citenamefont {Tews},
  \citenamefont {Gandolfi}, \citenamefont {Gezerlis},\ and\ \citenamefont
  {Schwenk}}]{tews2016}%
  \BibitemOpen
  \bibfield  {author} {\bibinfo {author} {\bibfnamefont {I.}~\bibnamefont
  {Tews}}, \bibinfo {author} {\bibfnamefont {S.}~\bibnamefont {Gandolfi}},
  \bibinfo {author} {\bibfnamefont {A.}~\bibnamefont {Gezerlis}}, \ and\
  \bibinfo {author} {\bibfnamefont {A.}~\bibnamefont {Schwenk}},\ }\href
  {\doibase 10.1103/PhysRevC.93.024305} {\bibfield  {journal} {\bibinfo
  {journal} {Phys. Rev. C}\ }\textbf {\bibinfo {volume} {93}},\ \bibinfo
  {pages} {024305} (\bibinfo {year} {2016})}\BibitemShut {NoStop}%
\bibitem [{\citenamefont {Lynn}\ \emph {et~al.}(2016)\citenamefont {Lynn},
  \citenamefont {Tews}, \citenamefont {Carlson}, \citenamefont {Gandolfi},
  \citenamefont {Gezerlis}, \citenamefont {Schmidt},\ and\ \citenamefont
  {Schwenk}}]{lynn2016}%
  \BibitemOpen
  \bibfield  {author} {\bibinfo {author} {\bibfnamefont {J.~E.}\ \bibnamefont
  {Lynn}}, \bibinfo {author} {\bibfnamefont {I.}~\bibnamefont {Tews}}, \bibinfo
  {author} {\bibfnamefont {J.}~\bibnamefont {Carlson}}, \bibinfo {author}
  {\bibfnamefont {S.}~\bibnamefont {Gandolfi}}, \bibinfo {author}
  {\bibfnamefont {A.}~\bibnamefont {Gezerlis}}, \bibinfo {author}
  {\bibfnamefont {K.~E.}\ \bibnamefont {Schmidt}}, \ and\ \bibinfo {author}
  {\bibfnamefont {A.}~\bibnamefont {Schwenk}},\ }\href {\doibase
  10.1103/PhysRevLett.116.062501} {\bibfield  {journal} {\bibinfo  {journal}
  {Phys. Rev. Lett.}\ }\textbf {\bibinfo {volume} {116}},\ \bibinfo {pages}
  {062501} (\bibinfo {year} {2016})}\BibitemShut {NoStop}%
\bibitem [{\citenamefont {Gandolfi}\ \emph {et~al.}(2011)\citenamefont
  {Gandolfi}, \citenamefont {Carlson},\ and\ \citenamefont
  {Pieper}}]{gandolfi2011}%
  \BibitemOpen
  \bibfield  {author} {\bibinfo {author} {\bibfnamefont {S.}~\bibnamefont
  {Gandolfi}}, \bibinfo {author} {\bibfnamefont {J.}~\bibnamefont {Carlson}}, \
  and\ \bibinfo {author} {\bibfnamefont {S.~C.}\ \bibnamefont {Pieper}},\
  }\href {\doibase 10.1103/PhysRevLett.106.012501} {\bibfield  {journal}
  {\bibinfo  {journal} {Phys. Rev. Lett.}\ }\textbf {\bibinfo {volume} {106}},\
  \bibinfo {pages} {012501} (\bibinfo {year} {2011})}\BibitemShut {NoStop}%
\bibitem [{\citenamefont {Bohr}\ and\ \citenamefont
  {Mottelson}(1998)}]{BohrMottelson}%
  \BibitemOpen
  \bibfield  {author} {\bibinfo {author} {\bibfnamefont {A.}~\bibnamefont
  {Bohr}}\ and\ \bibinfo {author} {\bibfnamefont {B.~R.}\ \bibnamefont
  {Mottelson}},\ }\href@noop {} {\emph {\bibinfo {title} {Nuclear
  Structure}}},\ Vol.~\bibinfo {volume} {I}\ (\bibinfo  {publisher} {World
  Scientific},\ \bibinfo {address} {Singapore},\ \bibinfo {year}
  {1998})\BibitemShut {NoStop}%
\bibitem [{\citenamefont {Schmidt}\ and\ \citenamefont
  {Fantoni}(1999)}]{schmidt1999}%
  \BibitemOpen
  \bibfield  {author} {\bibinfo {author} {\bibfnamefont {K.~E.}\ \bibnamefont
  {Schmidt}}\ and\ \bibinfo {author} {\bibfnamefont {S.}~\bibnamefont
  {Fantoni}},\ }\href {\doibase 10.1016/S0370-2693(98)01522-6} {\bibfield
  {journal} {\bibinfo  {journal} {Phys. Lett. B}\ }\textbf {\bibinfo {volume}
  {446}},\ \bibinfo {pages} {99} (\bibinfo {year} {1999})}\BibitemShut
  {NoStop}%
\bibitem [{\citenamefont {Carlson}\ \emph {et~al.}(2015)\citenamefont
  {Carlson}, \citenamefont {Gandolfi}, \citenamefont {Pederiva}, \citenamefont
  {Pieper}, \citenamefont {Schiavilla}, \citenamefont {Schmidt},\ and\
  \citenamefont {Wiringa}}]{carlson2015}%
  \BibitemOpen
  \bibfield  {author} {\bibinfo {author} {\bibfnamefont {J.}~\bibnamefont
  {Carlson}}, \bibinfo {author} {\bibfnamefont {S.}~\bibnamefont {Gandolfi}},
  \bibinfo {author} {\bibfnamefont {F.}~\bibnamefont {Pederiva}}, \bibinfo
  {author} {\bibfnamefont {S.~C.}\ \bibnamefont {Pieper}}, \bibinfo {author}
  {\bibfnamefont {R.}~\bibnamefont {Schiavilla}}, \bibinfo {author}
  {\bibfnamefont {K.~E.}\ \bibnamefont {Schmidt}}, \ and\ \bibinfo {author}
  {\bibfnamefont {R.~B.}\ \bibnamefont {Wiringa}},\ }\href {\doibase
  10.1103/RevModPhys.87.1067} {\bibfield  {journal} {\bibinfo  {journal} {Rev.
  Mod. Phys.}\ }\textbf {\bibinfo {volume} {87}},\ \bibinfo {pages} {1067}
  (\bibinfo {year} {2015})}\BibitemShut {NoStop}%
\bibitem [{\citenamefont {Pethick}\ \emph {et~al.}(1995)\citenamefont
  {Pethick}, \citenamefont {Ravenhall},\ and\ \citenamefont
  {Lorenz}}]{pethick1995}%
  \BibitemOpen
  \bibfield  {author} {\bibinfo {author} {\bibfnamefont {C.~J.}\ \bibnamefont
  {Pethick}}, \bibinfo {author} {\bibfnamefont {D.~G.}\ \bibnamefont
  {Ravenhall}}, \ and\ \bibinfo {author} {\bibfnamefont {C.~P.}\ \bibnamefont
  {Lorenz}},\ }\href {\doibase 10.1016/0375-9474(94)00506-I} {\bibfield
  {journal} {\bibinfo  {journal} {Nucl. Phys. A}\ }\textbf {\bibinfo {volume}
  {584}},\ \bibinfo {pages} {675} (\bibinfo {year} {1995})}\BibitemShut
  {NoStop}%
\bibitem [{\citenamefont {Sorella}(2001)}]{sorella2001}%
  \BibitemOpen
  \bibfield  {author} {\bibinfo {author} {\bibfnamefont {S.}~\bibnamefont
  {Sorella}},\ }\href {\doibase 10.1103/PhysRevB.64.024512} {\bibfield
  {journal} {\bibinfo  {journal} {Phys. Rev. B}\ }\textbf {\bibinfo {volume}
  {64}},\ \bibinfo {pages} {024512} (\bibinfo {year} {2001})}\BibitemShut
  {NoStop}%
\bibitem [{\citenamefont {Epelbaum}\ \emph {et~al.}(2015)\citenamefont
  {Epelbaum}, \citenamefont {Krebs},\ and\ \citenamefont
  {Mei\ss{}ner}}]{epelbaum2015}%
  \BibitemOpen
  \bibfield  {author} {\bibinfo {author} {\bibfnamefont {E.}~\bibnamefont
  {Epelbaum}}, \bibinfo {author} {\bibfnamefont {H.}~\bibnamefont {Krebs}}, \
  and\ \bibinfo {author} {\bibfnamefont {U.-G.}\ \bibnamefont {Mei\ss{}ner}},\
  }\href {\doibase 10.1140/epja/i2015-15053-8} {\bibfield  {journal} {\bibinfo
  {journal} {Eur. Phys. J. A}\ }\textbf {\bibinfo {volume} {51}},\ \bibinfo
  {pages} {53} (\bibinfo {year} {2015})}\BibitemShut {NoStop}%
\bibitem [{\citenamefont {Fossez}\ \emph {et~al.}()\citenamefont {Fossez},
  \citenamefont {Rotureau}, \citenamefont {Michel},\ and\ \citenamefont
  {{P\l{}oszajczak}}}]{fossez2016}%
  \BibitemOpen
  \bibfield  {author} {\bibinfo {author} {\bibfnamefont {K.}~\bibnamefont
  {Fossez}}, \bibinfo {author} {\bibfnamefont {J.}~\bibnamefont {Rotureau}},
  \bibinfo {author} {\bibfnamefont {N.}~\bibnamefont {Michel}}, \ and\ \bibinfo
  {author} {\bibfnamefont {M.}~\bibnamefont {{P\l{}oszajczak}}},\ }\href@noop
  {} {\ }\Eprint {http://arxiv.org/abs/1612.01483} {arXiv:1612.01483}
  \BibitemShut {NoStop}%
\bibitem [{\citenamefont {{Chang}}\ \emph {et~al.}(2004)\citenamefont
  {{Chang}}, \citenamefont {{Morales}}, \citenamefont {{Pandharipande}},
  \citenamefont {{Ravenhall}}, \citenamefont {{Carlson}}, \citenamefont
  {{Pieper}}, \citenamefont {{Wiringa}},\ and\ \citenamefont
  {{Schmidt}}}]{Chang:2004}%
  \BibitemOpen
  \bibfield  {author} {\bibinfo {author} {\bibfnamefont {S.-Y.}\ \bibnamefont
  {{Chang}}}, \bibinfo {author} {\bibfnamefont {J.}~\bibnamefont {{Morales}}},
  \bibinfo {author} {\bibfnamefont {V.~R.}\ \bibnamefont {{Pandharipande}}},
  \bibinfo {author} {\bibfnamefont {D.~G.}\ \bibnamefont {{Ravenhall}}},
  \bibinfo {author} {\bibfnamefont {J.}~\bibnamefont {{Carlson}}}, \bibinfo
  {author} {\bibfnamefont {S.~C.}\ \bibnamefont {{Pieper}}}, \bibinfo {author}
  {\bibfnamefont {R.~B.}\ \bibnamefont {{Wiringa}}}, \ and\ \bibinfo {author}
  {\bibfnamefont {K.~E.}\ \bibnamefont {{Schmidt}}},\ }\href {\doibase
  10.1016/j.nuclphysa.2004.09.119} {\bibfield  {journal} {\bibinfo  {journal}
  {Nucl. Phys. A}\ }\textbf {\bibinfo {volume} {746}},\ \bibinfo {pages} {215}
  (\bibinfo {year} {2004})}\BibitemShut {NoStop}%
\bibitem [{\citenamefont {Fetter}\ and\ \citenamefont
  {Walecka}(1971)}]{FW:1971}%
  \BibitemOpen
  \bibfield  {author} {\bibinfo {author} {\bibfnamefont {A.~L.}\ \bibnamefont
  {Fetter}}\ and\ \bibinfo {author} {\bibfnamefont {J.~D.}\ \bibnamefont
  {Walecka}},\ }\href@noop {} {\emph {\bibinfo {title} {Quantum Theory of
  Many-Particle Systems}}}\ (\bibinfo  {publisher} {McGraw-Hill},\ \bibinfo
  {address} {New York},\ \bibinfo {year} {1971})\BibitemShut {NoStop}%
\bibitem [{\citenamefont {De{M}arco}\ and\ \citenamefont
  {Jin}(1999)}]{deMarco:1999}%
  \BibitemOpen
  \bibfield  {author} {\bibinfo {author} {\bibfnamefont {B.}~\bibnamefont
  {De{M}arco}}\ and\ \bibinfo {author} {\bibfnamefont {D.~S.}\ \bibnamefont
  {Jin}},\ }\href {\doibase 10.1126/science.285.5434.1703} {\bibfield
  {journal} {\bibinfo  {journal} {Science}\ }\textbf {\bibinfo {volume}
  {285}},\ \bibinfo {pages} {1703} (\bibinfo {year} {1999})}\BibitemShut
  {NoStop}%
\bibitem [{\citenamefont {Regal}\ and\ \citenamefont {Jin}(2003)}]{Regal:2003}%
  \BibitemOpen
  \bibfield  {author} {\bibinfo {author} {\bibfnamefont {C.~A.}\ \bibnamefont
  {Regal}}\ and\ \bibinfo {author} {\bibfnamefont {D.~S.}\ \bibnamefont
  {Jin}},\ }\href {\doibase 10.1103/PhysRevLett.90.230404} {\bibfield
  {journal} {\bibinfo  {journal} {Phys. Rev. Lett.}\ }\textbf {\bibinfo
  {volume} {90}},\ \bibinfo {pages} {230404} (\bibinfo {year}
  {2003})}\BibitemShut {NoStop}%
\bibitem [{\citenamefont {Wenz}\ \emph {et~al.}(2013)\citenamefont {Wenz},
  \citenamefont {Z{\"u}rn}, \citenamefont {Murmann}, \citenamefont {Brouzos},
  \citenamefont {Lompe},\ and\ \citenamefont {Jochim}}]{Zuern:2013}%
  \BibitemOpen
  \bibfield  {author} {\bibinfo {author} {\bibfnamefont {A.~N.}\ \bibnamefont
  {Wenz}}, \bibinfo {author} {\bibfnamefont {G.}~\bibnamefont {Z{\"u}rn}},
  \bibinfo {author} {\bibfnamefont {S.}~\bibnamefont {Murmann}}, \bibinfo
  {author} {\bibfnamefont {I.}~\bibnamefont {Brouzos}}, \bibinfo {author}
  {\bibfnamefont {T.}~\bibnamefont {Lompe}}, \ and\ \bibinfo {author}
  {\bibfnamefont {S.}~\bibnamefont {Jochim}},\ }\href {\doibase
  10.1126/science.1240516} {\bibfield  {journal} {\bibinfo  {journal}
  {Science}\ }\textbf {\bibinfo {volume} {342}},\ \bibinfo {pages} {457}
  (\bibinfo {year} {2013})}\BibitemShut {NoStop}%
\end{thebibliography}%

\end{document}